\documentclass[aps,prd,floatfix,superscriptaddress,balancelastpage,nofootinbib,amsmath,showpacs,twocolumn]{revtex4-1}
\usepackage{graphicx,epstopdf,epsfig,bm,dcolumn,natbib,amsmath,amssymb,multirow}

\newcommand{\beq}{\begin{equation}} \newcommand{\eeq}{\end{equation}}
\newcommand{\bea}{\begin{eqnarray}} \newcommand{\eea}{\end{eqnarray}}
           
\newcommand{\pL}{\left(} \newcommand{\pR}{\right)} \newcommand{\bL}{\left[} \newcommand{\bR}{\right]}

\def\lsim{\mathrel{\raise.3ex\hbox{$<$\kern-.75em\lower1ex\hbox{$\sim$}}}}
\def\gsim{\mathrel{\raise.3ex\hbox{$>$\kern-.75em\lower1ex\hbox{$\sim$}}}}

\begin{document}

\title{$Z'$ Mediated Dark Matter Models for the Galactic Center Gamma-Ray Excess}

\author{Dan Hooper}
\affiliation{Center for Particle Astrophysics, Fermi National Accelerator Laboratory, Batavia, IL 60510}
\affiliation{Department of Astronomy and Astrophysics, University of Chicago, Chicago, IL 60637}

\begin{abstract}

With the goal of generating the Galactic Center gamma-ray excess, we revisit models in which the dark matter interacts with the Standard Model through the exchange of a new neutral gauge boson, $Z'$. We find several scenarios that can account for this signal, while respecting all existing constraints from colliders and direct detection experiments. In such models, the $Z'$ either 1) couples axially to light quarks and is leptophobic, 2) couples dominantly to the third generation, or 3) is near resonance, $m_{Z'} \approx 2 m_{\rm DM}$. We identify an example of an anomaly-free $U(1)'$ that leads to an axial and leptophobic $Z'$. Many of the models presented here are within the reach of near future direct detection experiments, such as LUX and XENON1T.

\end{abstract}

\pacs{95.35.+d, 14.70.Pw, 95.85.Pw; FERMILAB-PUB-14-473-A} 

\maketitle

\section{Introduction}

The excess of gamma-rays observed from the region surrounding the Galactic Center~\cite{Goodenough:2009gk,Hooper:2010mq,Boyarsky:2010dr,Hooper:2011ti,Abazajian:2012pn,Gordon:2013vta,Hooper:2013rwa,Huang:2013pda,Abazajian:2014fta,Daylan:2014rsa,Calore:2014xka,fermigc} exhibits a number of characteristics which favor an interpretation in terms of annihilating dark matter. In particular, the spectrum, spatial morphology, and overall normalization of this signal are each consistent with that predicted from annihilating weakly interacting massive particles (WIMPs).  And although numerous particle physics models have been proposed to account for this signal, the overwhelming majority fall within four phenomenological categories:
\begin{itemize}
\item{Models in which the dark matter annihilates through the $s$-channel exchange of a spin-zero mediator with pseudoscalar couplings~\cite{Ipek:2014gua,Boehm:2014hva,Berlin:2014tja,Izaguirre:2014vva,Cheung:2014lqa,Cerdeno:2014cda,Alves:2014yha}.}
\item{Models in which the dark matter annihilates to $b\bar{b}$ through the $t$-channel exchange of a colored and charged state with a TeV-scale mass~\cite{Agrawal:2014una,Berlin:2014tja}.}
\item{Models in which the dark matter annihilates to produce unstable particles with only very small couplings to the Standard Model (SM) ({\it ie}.~hidden sector models)~\cite{Hooper:2012cw,Berlin:2014pya,Ko:2014gha,Boehm:2014bia,Abdullah:2014lla,Martin:2014sxa,Cline:2014dwa,Ko:2014loa,Freytsis:2014sua}.}
\item{Models in which the dark matter annihilates through the $s$-channel exchange of a new neutral gauge boson, $Z'$~\cite{Berlin:2014tja,Izaguirre:2014vva,Alves:2014yha}.}
\end{itemize}
In this article, we revisit the last of these four possibilities, considering the constraints on such scenarios from direct detection experiments, as well as from the Large Hadron Collider (LHC) and other accelerator experiments. With these constraints in mind, we consider options for model building, identifying several phenomenologically viable and anomaly-free models that are capable of explaining the Galactic Center excess.

The existence of gauge symmetries beyond those of the Standard Model are a feature common to a wide variety of new physics scenarios (for a review, see Ref.~\cite{Langacker:2008yv}). In particular, new broken abelian $U(1)$ gauge symmetries and the $Z'$ bosons that accompany them are predicted by many Grand Unified Theories (GUTs), including those based on the $SO(10)$ or $E_6$ groups~\cite{London:1986dk,Hewett:1988xc}. Additional massive gauge bosons also appear as a generic element of string inspired phenomenological models~\cite{Braun:2005bw,Cleaver:1998gc,Coriano:2007ba,Faraggi:1990ita,Giedt:2000bi,Lebedev:2007hv,Anastasopoulos:2006cz,Faraggi:1991mu,Cvetic:2001nr}, as well as within the context of little Higgs theories~\cite{ArkaniHamed:2001is,ArkaniHamed:2002qx,Han:2003wu,Perelstein:2005ka}, dynamical symmetry breaking scenarios~\cite{Hill:2002ap,Chivukula:2003wj,Chivukula:2002ry}, models with extra spatial dimensions~\cite{Agashe:2003zs,Agashe:2007ki,Carena:2003fx,Hewett:2002fe}, and many other popular extensions of the Standard Model~\cite{ArkaniHamed:2001nc,Cvetic:1997ky,Langacker:1999hs}. It has also been long appreciated that the dark matter might interact predominantly through its couplings to a $Z'$, leading to a number of implications for direct and indirect detection~\cite{Agashe:2007jb,Agashe:2004bm,Agashe:2004ci,Lee:2007mt,Pospelov:2007mp,Buckley:2011mm,Buckley:2011vs,Belanger:2007dx,Hur:2007ur,Lebedev:2014bba,Arcadi:2013qia,Dudas:2013sia,Chu:2013jja,Mambrini:2011dw,Dudas:2009uq}.

In this paper, we revisit $Z'$ mediated dark matter scenarios within the context of the Galactic Center gamma-ray excess.  After reviewing the characteristics of the excess in Sec.~\ref{excess}, we consider in Sec.~\ref{simplified} a number of simplified models, calculating the thermal relic abundance, low-velocity annihilation cross section, and elastic scattering cross section of the dark matter in a variety of $Z'$ mediated scenarios. In Sec.~\ref{collidersec}, we review the constraints from the LHC (as well as from the Tevatron, UA2, and LEP) on models with a $Z'$. With these collider and direct detection constraints in mind, we turn in Sec.~\ref{anomaliessec} toward model building, identifying a number of anomaly-free $U(1)'$s that are potentially capable of explaining the Galactic Center excess. In Secs.~\ref{future} and~\ref{conclusion}, we discuss the prospects for testing this collection of models with near future experiments and summarize our results and conclusions.

\section{The Galactic Center Gamma-Ray Excess}
\label{excess}

Over the past several years, a gamma-ray excess from the inner kiloparsecs surrounding the Galactic Center has been identified from within the data of the Fermi Gamma-Ray Space Telescope~\cite{Goodenough:2009gk,Hooper:2010mq,Boyarsky:2010dr,Hooper:2011ti,Abazajian:2012pn,Gordon:2013vta,Hooper:2013rwa,Huang:2013pda,Abazajian:2014fta,Daylan:2014rsa,Calore:2014xka,fermigc}. Many of characteristics of this signal are consistent with those predicted from the annihilation of dark matter particles. In particular, the signal is distributed with approximate spherical symmetry about the Galactic Center, with an angular distribution that corresponds to a dark matter density profile of $\rho_{\rm DM} \propto r^{-1.2}$, in agreement with expectations from hydrodynamical simulations of Milky Way-like galaxies~\cite{DiCintio:2014xia,DiCintio:2013qxa}. The spectral shape of the observed excess is also in good agreement with that expected from annihilating dark matter. Considering dark matter particles that annihilate predominately to $b\bar{b}$, for example, the observed spectrum favors a mass in the range of 31-40 GeV~\cite{Daylan:2014rsa} (or 43-55 GeV~\cite{Calore:2014xka}, 35-43 GeV~\cite{Abazajian:2014fta}, as found by the authors of other recent analyses; see also Ref.~\cite{Agrawal:2014oha}). Annihilations to lighter quarks (or to combinations of quark species) can also provide a good fit, although for somewhat lower values of the dark matter's mass~\cite{Daylan:2014rsa}.

Recently, the authors of Ref.~\cite{Calore:2014xka} studied how systematic uncertainties related to the modeling of diffuse astrophysical emission processes (pion production, inverse Compton scattering, and bremsstrahlung) might impact the spectrum and angular distribution of the Galactic Center gamma-ray excess.  Despite considering a wide range of assumptions regarding the distribution of cosmic ray sources, gas, radiation, and magnetic fields, as well as for the diffusion, convection and reacceleration of cosmic rays, it was found that the excess persists for all such variations. Furthermore, the spectral and morphological characteristics of the excess were found to be quite stable across this range of astrophysical assumptions. Similar conclusions were also reached in Ref.~\cite{Zhou:2014lva}.

To date, two classes of astrophysical explanations for the Galactic Center excess have been proposed. The first of these is that a population of thousands of unresolved millisecond pulsars might generate the observed signal. Millisecond pulsars were originally suggested as a possible source for the excess due to the fact that the shape of the gamma-ray spectrum observed from such sources is similar to that predicted from the annihilations of $\sim$30 GeV dark matter particles~\cite{Hooper:2010mq,Abazajian:2010zy} (and also as presciently noted in Ref.~\cite{Baltz:2006sv}). In light of recent analyses, which revealed the Galactic Center excess to be spatially extended to a radius of at least $\sim$$10^{\circ}$~\cite{Hooper:2013rwa,Daylan:2014rsa,Calore:2014xka}, such sources no longer appear to be a tenable explanation. In particular, if enough millisecond pulsars were present within the inner kiloparsecs of the galaxy to account for the observed emission, Fermi should have detected and resolved a large number of such bright objects ($\sim$60 with $L_{\gamma}>10^{35}$ erg/s)~\cite{Cholis:2014lta,Hooper:2013nhl,Cholis:2014noa}, whereas none have been observed. Only if the luminosity function of the millisecond pulsar population in the region surrounding the Galactic Center is significantly different from those observed elsewhere~\cite{Yuan:2014rca,Petrovic:2014xra} could this constraint be evaded. Furthermore, a comparison of the numbers of luminous low mass X-ray binaries in globular clusters and in the Galactic Bulge suggests that the gamma-ray emission from millisecond pulsars is likely to constitute a flux that is only a few percent as bright as that of the gamma-ray excess~\cite{Cholis:2014lta}. 

The second proposed astrophysical explanation for the excess is a recent outburst of cosmic rays from the Galactic Center, generating the gamma-ray signal via either hadronic (pion production)~\cite{Carlson:2014cwa} or leptonic (inverse Compton scattering)~\cite{Petrovic:2014uda} processes. The hadronic case, however, inevitably leads to a gamma-ray signal that is correlated with the distribution of gas and that is extended along the Galactic Plane~\cite{Carlson:2014cwa}, highly incompatible with the observed characteristics of the excess. The leptonic case, in contrast, could be more smoothly distributed and spherically symmetric. It is difficult, however, for inverse Compton scattering to simultaneously generate a spectrum and spatial morphology that is compatible with the observed emission~\cite{Petrovic:2014uda}, in particular given the apparent consistency of the spectral shape across different regions of the Inner Galaxy~\cite{Calore:2014xka}. 


Interpretations of the Galactic Center excess in terms of annihilating dark matter often assume that the dark matter is distributed according to a generalized NFW profile:
\begin{equation}
\rho( r)= \frac{\rho_0}{(r/R_s)^{\gamma} [1+(r/R_s)]^{3-\gamma}}, 
\end{equation}
where $\gamma$ is the profile's inner slope and $R_s$ is the scale radius. For example, Ref.~\cite{Daylan:2014rsa} adopts a scale radius of $R_s=20$ kpc and a value of $\rho_0$ such that $\rho_{\rm local}=0.3$ GeV/cm$^3$, and then finds that $\gamma=1.23-1.29$ provides the best fit to the data. For annihilations to $b\bar{b}$, and under these assumptions, the annihilation cross section required to fit the excess was found to be $\sigma v = (1.7-2.3) \times 10^{-26}$ cm$^3/$s. If we allow the local density and scale radius of the profile to vary within the range allowed by dynamical constraints ($\rho_{\rm local}=$0.24-0.46 GeV/cm$^3$~\cite{Iocco:2011jz}, $R_s=$8-35 kpc~\cite{Catena:2009mf,Bergstrom:1997fj,Iocco:2011jz}) while requiring the slope of the halo profile at $r=0.3$ kpc to remain fixed to the best fit value, this allows for a significantly wider range of values for the annihilation cross section, $\sigma v \simeq (0.4-5) \times 10^{-26}$ cm$^3/$s. If one were to allow for departures from the simple parameterization of the generalized NFW profile, an even larger range of cross sections would be found to be compatible with the observed signal. 

The Fermi Collaboration has recently presented results from a stacked analysis of dwarf spheroidal galaxies, finding that $\sigma v \lsim 1.4 \times 10^{-26}$ cm$^3$/s for $m_{\rm DM}=35$ GeV and annihilations to $b\bar{b}$ (assuming that dark matter is distributed in dwarf spheroidals according to an NFW profile)~\cite{fermidwarf} (see also Ref.~\cite{Geringer-Sameth:2014qqa}). While this new limit does appear to exclude the largest values of the cross section that are compatible with the Galactic Center excess, it is not yet in tension with a dark matter interpretation of this signal. This constraint may suggest, however, a low-velocity dark matter annihilation cross section that is somewhat lower than its value at the temperature of freeze-out. Such a difference could arise from a $p$-wave contribution to the amplitude for dark matter annihilation, or from annihilations taking place near a resonance. As we will show, such phenomenology can arise in a variety of $Z'$ mediated scenarios.

\section{Simplified Models for $Z'$ Mediated Dark Matter}
\label{simplified}

In this section, we take a model-independent approach to $Z'$ mediated dark matter, following that taken in Ref.~\cite{Berlin:2014tja}. In later sections, we will consider how the phenomenological features discussed in this section might arise within the context of a larger and more complete theory. 

At first glance, it may appear that a wide variety of $Z'$ mediated scenarios could potentially account for the observed gamma-ray excess. After all accelerator and direct detection constraints are taken into account, however, we find that relatively few phenomenological options are presently viable. For example, unless the dark matter is a fermion, $Z'$ mediated annihilation diagrams are suppressed by powers of velocity ($p$-wave suppressed), and thus are unable to generate the observed flux of gamma-rays. Furthermore, a $Z'$ with vector couplings to light quarks will generate an elastic scattering cross section with nuclei that is strongly ruled out by direct dark matter searches. There are, however, more promising scenarios, such as dark matter in the form of a Dirac or Majorana fermion that annihilates through the exchange of a $Z'$ with purely axial couplings to quarks. Alternatively, if the mediating $Z'$ couples only to third generation fermions (or otherwise has suppressed couplings to light quarks and electrons), we can identify a number of scenarios that are consistent with the constraints of accelerators and direct searches. In the following subsections, we will discuss the annihilation, relic abundance, and elastic scattering of dark matter in each of the aforementioned cases.

\subsection{Dirac Dark Matter With An Axial $Z'$}
\label{axial}

Consider a dark matter particle that is a Dirac fermion, $\chi$, coupled to a spin-$1$ mediator with axial couplings to SM fermions, $Z'_{\mu}$:
\begin{equation}
\mathcal{L} \supset \left[ \bar{\chi} \gamma^\mu ( g_{\chi  v } + g_{\chi a} \gamma^5 ) \chi + g_f \bar{f} \gamma^\mu \gamma^5  f \right] Z'_\mu. \\
\end{equation}
Note that we deliberately do not include a $\bar{f} \gamma^{\mu} f Z'_{\mu}$ term in order to avoid generating a large vector amplitude for elastic scattering. The annihilation cross section in this case is given by:
\begin{widetext}
\begin{eqnarray}
\label{one}
\sigma &=& \sum_{f} \frac{n_c }{12 \pi  s \bL \left(s-m_{Z'} ^2\right)^2  + m_{Z'}^2 \Gamma_{Z'}^2 \bR } \sqrt{\frac{1-4 m_f^2/s}{1-4 m_{\chi }^2/s}} \\
&\times& g_{f}^2 \Bigg[g_{\chi a}^2 \bigg\{4 m_{\chi }^2 \bigg[m_f^2 \pL 7- \frac{6s}{m_{Z'}^2}+\frac{3s^2}{m_{Z'}^4} \pR-s \bigg]+s  \left(s-4 m_f^2\right)\bigg\} +g_{\chi v}^2  (s-4 m_f^2) (2 m_{\chi}^2+s)  \Bigg], \label{sigsdirV} \nonumber
\end{eqnarray}
where $n_c=3$ (1) for annihilations to quarks (leptons), $\sqrt{s}$ is the center-of-mass energy of the collision, and $\Gamma_{Z'}$ is width of the $Z'$:
\beq
\Gamma_{Z'} \equiv \sum_f \Gamma(V \to f \bar f) =  \sum_f \frac{n_c m_{Z'} }{24 \pi}  \sqrt{1-\frac{4 m_f^2}{m_{Z'}^2}} \bL g_{fa}^2 \left(1-\frac{4 m_f^2}{m_{Z'}^2} \right)+g_{fv}^2 \left(1+ 2 \frac{m_f^2}{m_{Z'}^2} \pR \bR.
\eeq
Although we will use the full expression given in Eq.~\ref{one} for the calculation of the relic abundance, it is illustrative to expand the cross section in powers of velocity.  Presented this way, the cross section (for each SM fermion species) is given by:
\begin{eqnarray}
\label{two}
\sigma v &\approx& \frac{n_c \sqrt{1-m_f^2/m_\chi^2}}{2 \pi m_{Z'}^4  \left(m_{Z'}^2-4 m_{\chi }^2\right)^2} \,\, g_{f}^2 \Bigg[  m_f^2 g_{\chi a}^2 \left(m_{Z'}^2-4 m_{\chi }^2\right)^2+2 g_{\chi v}^2 m_{Z'}^4 \left(m_{\chi }^2-m_f^2\right) \Bigg] \\
&-& \frac{n_c v^2}{48 \pi  m_{Z'}^4 m_{\chi }^2 \sqrt{1-m_f^2/m_{\chi }^2} \left(4 m_{\chi }^2-m_{Z'}^2\right)^3} 
\,\, g_{f}^2  \Bigg[ g_{\chi a}^2 \left(m_{Z'}^2-4 m_{\chi }^2\right) \Big(m_f^4 \left(-72 m_{Z'}^2 m_{\chi }^2+17 m_{Z'}^4+144 m_{\chi }^4\right) \nonumber \\
&+&m_f^2 \left(48 m_{Z'}^2 m_{\chi }^4-22 m_{Z'}^4 m_{\chi }^2-96 m_{\chi }^6\right)+8 m_{Z'}^4 m_{\chi }^4\Big)  
  - 2 g_{\chi v}^2 m_{Z'}^4 \left(m_f^2-m_{\chi }^2\right) \Big(4 m_{\chi }^2 \left(m_{Z'}^2-17
   m_f^2\right)+5 m_f^2 m_{Z'}^2+32 m_{\chi }^4\Big) \Bigg] \nonumber.
\end{eqnarray}
The first term of this expression generically dominates the annihilation cross section at thermal freeze-out, and provides an unsuppressed annihilation rate in the universe today. Also note that so long as $g_{\chi v}$ is not much smaller than $g_{\chi a}$, the annihilation will proceed approximately equally to all SM fermion final states, according to their couplings to the $Z'$. A somewhat counterintuitive aspect of this result is that for a purely axial coupling ($g_{\chi v}=0$), the low-velocity cross section does not exhibit a pole at $m_{Z'}=2m_{\chi}$. 

For this case of a $Z'$ with purely axial couplings to quarks, the dark matter interacts with nuclei only through spin-dependent operators:
\begin{equation}
\sigma \simeq  \frac{ \mu_{\chi N}^2 (4  g_{\chi a}^2+2 g_{\chi v}^2 v^2 \mu_{\chi N}^2/ \mu_{\chi n}^2)}{\pi m_{Z'}^4} J_N(J_N+1) \bigg[ \frac{\langle S_p \rangle}{J_N}  (g_{u} \Delta^{( p )}_u+ g_{d} \Delta^{( p )}_d + g_{s} \Delta^{( p )}_s ) +  \frac{\langle S_n \rangle}{J_N}  (g_{u} \Delta^{( n )}_u+ g_{d} \Delta^{( n )}_d + g_{s} \Delta^{( n )}_s ) \bigg]^2,
\label{dirSDa} 
\end{equation}
\end{widetext}
where $J_N$ is the spin of the target nucleus. We take the standard values to describe the nuclear quark content, $\Delta_ u^{(p)}=\Delta_ d^{(n)}=0.84$, $\Delta_ u^{(n)}=\Delta_ d^{(p)}=-0.43$, and $\Delta_ s^{(p)}=\Delta_ s^{(n)}=-0.09$~\cite{pdg}. Note that the vector coupling, $g_{\chi v}$, leads to an elastic scattering cross section that is suppressed by two powers of the velocity, resulting from the fact that the momentum transfer does not sum coherently in the case of this operator.



\begin{figure}
\includegraphics[width=3.3in]{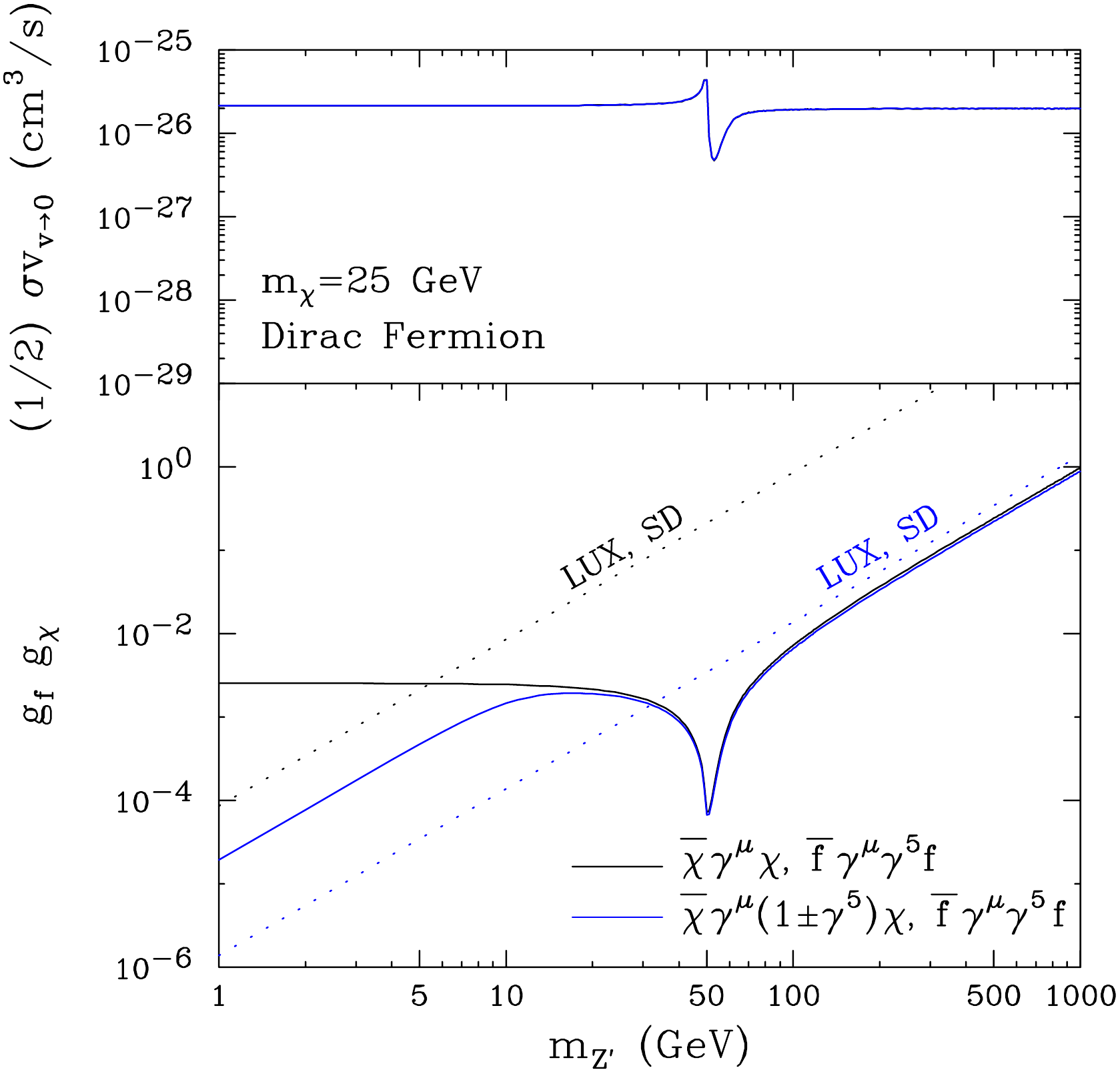} 
\caption{Results for dark matter in the form of a Dirac fermion, annihilating through the exchange of a $Z'$ with only axial couplings to Standard Model fermions (assumed to be universal/democratic). In the lower frame, the solid lines denote the product of the couplings $g_f$, $g_{\chi}$ needed to generate a thermal relic abundance equal to the observed cosmological dark matter density. The dotted lines represent the current constraints from direct detection experiments~\cite{Akerib:2013tjd}. In the upper frame, we plot the low-velocity annihilation cross section (as relevant for indirect detection). In each frame, the black and blue contours correspond to the case of dark matter with purely vector couplings to the $Z'$, and that of equal vector and axial couplings.}
\label{diracaxial}
\end{figure}

In the lower portion of Fig.~\ref{diracaxial}, we plot the couplings required to generate a thermal relic abundance equal to the measured cosmological dark matter density (as calculated following Ref.~\cite{Griest:1990kh}), for the cases of a purely vector coupling to dark matter (solid black) and equal vector and axial couplings to dark matter (solid blue), assuming democratic couplings to all SM fermions. In the calculation of the thermal relic abundance, we take $\Gamma_{Z'}=1$ GeV (the width cannot be calculated without specifying the value of  $g_f$). In the upper portion of the same figure, the low velocity annihilation cross section for this choice of couplings is shown.  The factor of $1/2$ included in this quantity accounts for the fact that the dark matter is Dirac, and thus the annihilation rate in the Galactic Center is suppressed relative to the Majorana case. In this scenario, the cross section is dominated by the $s$-wave contribution, leading to an unsuppressed low-velocity cross section to heavy and light fermions alike. The results shown here are for the case of dark matter particle with a 25 GeV mass, which provides a good fit to the gamma-ray excess for annihilation channels that are approximately democratic. 

The dotted contours shown in the lower portion of Fig.~\ref{diracaxial} represent the constraints on this model from direct searches, as derived from the results of the LUX experiment~\cite{Akerib:2013tjd}.  If the $Z'$ interacts with the dark matter only through a vector coupling (dotted black), this model is likely to remain beyond the reach of direct detection experiments for the foreseeable future (unless the $Z'$ is very light). On the other hand, if the axial coupling is comparable to the vector coupling (dotted blue), direct detection experiments are expected to become sensitive to this scenario in the near future (with XENON1T, and possibly with additional data from LUX). Also note that if $m_{Z'} < m_{\chi}$, the dark matter will likely annihilate predominantly to $Z'$ pairs, depending on the ratio of the $Z'$'s couplings to SM fermions and dark matter; we do not consider this case further in this study.

\subsection{Majorana Dark Matter With An Axial or Axial-Vector $Z'$} 
\label{majoranasec}

\begin{figure}
\includegraphics[width=3.3in]{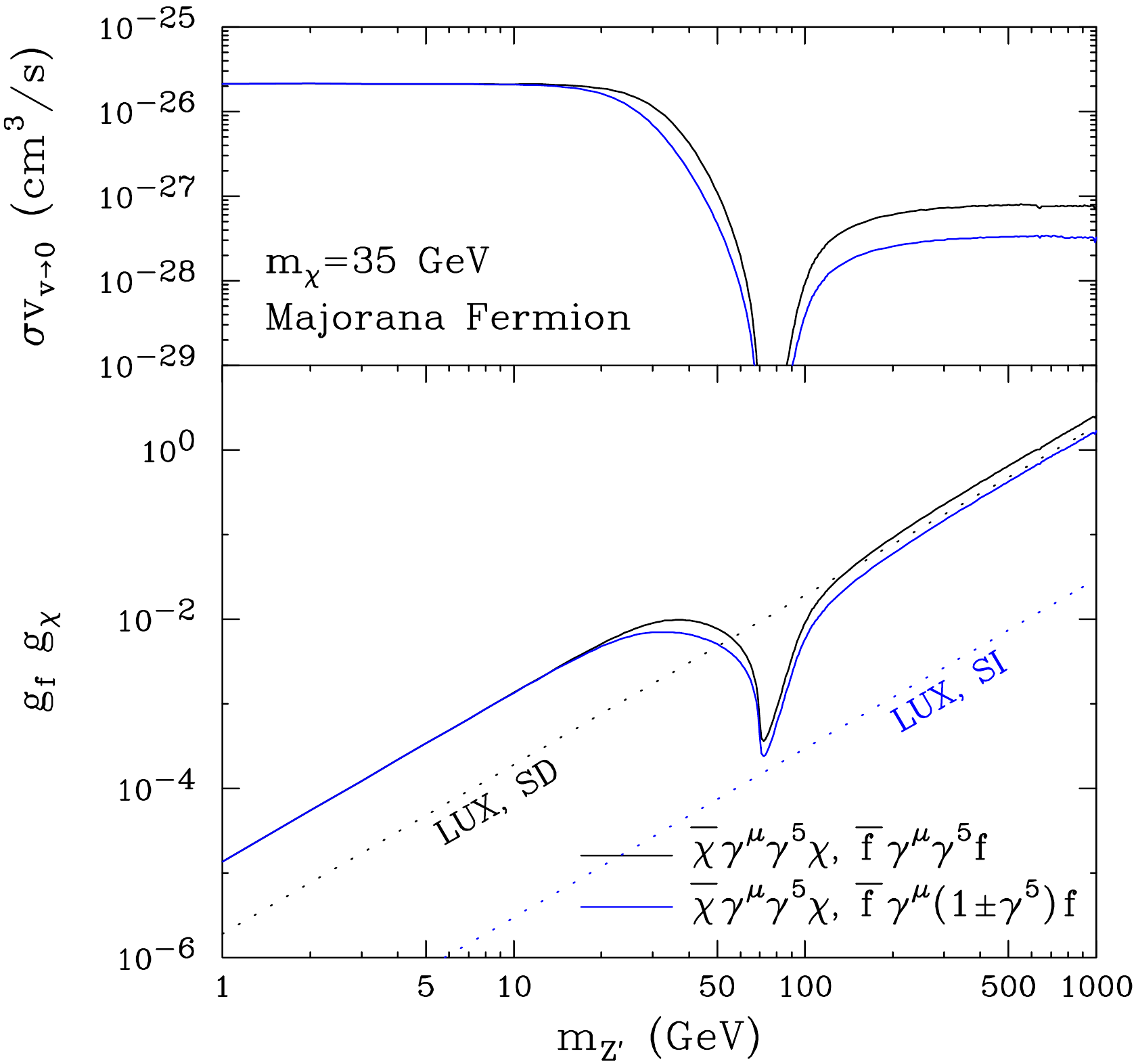} 
\caption{Results for dark matter in the form of a Majorana fermion, annihilating through the exchange of a $Z'$ with either axial (black) or axial-vector (blue) couplings to Standard Model fermions (assumed to be universal/democratic). In the lower frame, the solid lines denote the product of the couplings $g_f$, $g_{\chi}$ needed to generate a thermal relic abundance equal to the observed cosmological dark matter density. The dotted lines represent the current constraints from direct detection experiments~\cite{Akerib:2013tjd}. In the upper frame, we plot the low-velocity annihilation cross section (as relevant for indirect detection).}
\label{majorana}
\end{figure}

Next, we consider a Majorana dark matter particle, $\chi$, that interacts with SM fermions through a $Z'$:
\beq
\mathcal{L} \supset \left[ \frac{1}{2} g_{\chi} \bar{\chi} \gamma^\mu \gamma^5 \chi + \bar{f} \gamma^\mu ( g_{f v} + g_{f a} \gamma^5 ) f \right] Z'_\mu.
\eeq
Note that in the case of a Majorana fermion, the coupling to the $Z'$ is automatically purely-axial. The annihilation cross section in this case is given by:
\begin{widetext}
\begin{eqnarray}
\sigma &=&\sum_{f} \frac{n_c }{12 \pi  s \bL \left(s-m_{Z'} ^2\right)^2  + m_{Z'}^2 \Gamma_{Z'}^2 \bR } \sqrt{\frac{1-4 m_f^2/s}{1-4 m_{\chi }^2/s}} \\
&\times& g_{\chi}^2 \Bigg[g_{f a}^2\bigg\{ 4 m_{\chi }^2 \bigg[m_f^2 \pL 7- \frac{6s}{m_{Z'}^2}+\frac{3s^2}{m_{Z'}^4} \pR-s \bigg]+s  \left(s-4 m_f^2\right) \bigg\}+g_{fv}^2 (2 m_f^2+s) (s-4 m_{\chi }^2)  \Bigg]. \label{sigsdirV} \nonumber
\end{eqnarray}
Expanding in powers of velocity yields (for each SM fermion species):
\begin{eqnarray}
\sigma v  &\approx& \frac{n_c \, m_f^2  g_{fa}^2  g_{\chi }^2 \sqrt{1-m_f^2/m_\chi^2}}{2 \pi m_{Z'}^4} \\
&+& \frac{n_c v^2}{48 \pi  m_{Z'}^4 m_{\chi }^2 \sqrt{1-m_f^2/m_{\chi }^2} \left(4 m_{\chi }^2-m_{Z'}^2\right)^2} 
\,\,g^2_{\chi} \,\Bigg[ g_{fa}^2 \bigg\{ m_f^4 \left(-72 m_{Z'}^2 m_{\chi }^2+17 m_{Z'}^4+144 m_{\chi }^4\right) \nonumber \\
&+&m_f^2 \left(48 m_{Z'}^2 m_{\chi }^4-22 m_{Z'}^4 m_{\chi }^2-96 m_{\chi }^6\right)+8 m_{Z'}^4 m_{\chi }^4  \label{sigVsdirV}
\bigg\} \nonumber 
   -4g_{fv}^2 m_{Z'}^4 \left(m_f^2 m_{\chi
   }^2+m_f^4-2 m_{\chi }^4\right) \Bigg] \nonumber.
\end{eqnarray}
In contrast to the Dirac case, this cross section is dominated by the $v^2$ ($p$-wave) term at freeze-out (unless $m_{Z'} \lsim 2 m_{\chi}$). In other words, the $\sigma v \propto v^2$ term dominates the annihilation cross section for $v^2 \sim 0.1$, leading to an annihilation rate that is suppressed in the universe today (where $v^2 \sim 0$). This is illustrated in the upper portion of Fig.~\ref{majorana}, where the low-velocity cross section is shown to be well below the $s-$wave prediction for a thermal relic ($\sim2\times 10^{-26}$ cm$^3$/s). Note that this case is similar to that in which a Majorana dark matter particle annihilates through the exchange of the SM $Z$. As this leads to a low-velocity cross section on the order of $\sim$$10^{-28}$ cm$^3$/s (as seen from Fig.~\ref{majorana} for the case of $m_{Z'}=91.2$ GeV), dark matter annihilation through $Z$ exchange cannot be responsible for generating the Galactic Center gamma-ray excess. Once again, we note that the low-velocity cross section does not exhibit a pole at $m_{Z'} =2 m_{\chi}$.

The dark matter scatters with nuclei in this case through a combination of spin-dependent (axial) and spin-independent (vector) operators:
\begin{eqnarray}
\label{elasticmaj}
\sigma &\approx&  \frac{ \mu_{\chi N}^2 \,g_{\chi}^2}{\pi m_{Z'}^4} 4 J_N(J_N+1) \bigg[ \frac{\langle S_p \rangle}{J_N}  (g_{ua} \Delta^{( p )}_u+ g_{da} \Delta^{( p )}_d + g_{sa} \Delta^{( p )}_s ) +  \frac{\langle S_n \rangle}{J_N}  (g_{ua} \Delta^{( n )}_u+ g_{da} \Delta^{( n )}_d + g_{sa} \Delta^{( n )}_s ) \bigg]^2 \label{dirSDa}  \\
&+& \frac{\mu_{\chi N}^2 \, g_{\chi}^2}{\pi  m_{Z'}^4} \, \frac{2 \mu^4_{\chi N} v^2}{\mu_{\chi n}^2 m^2_{\chi}} \, \bigg[ Z \pL 2 g_{uv}+g_{dv} \pR + (A-Z) \pL g_{uv}+ 2 g_{dv} \pR \bigg]^2, \nonumber
\end{eqnarray}
where $Z$ and $A$ are the atomic number and atomic mass of the target, respectively.  The later term in this expression corresponds to an anapole moment interaction. For more detailed discussion, see Refs.~\cite{DelNobile:2014eta,Gresham:2014vja,Ho:2012bg}.

\end{widetext}

In Fig.~\ref{majorana}, we present the required couplings and low-velocity annihilation cross section in this scenario. Here, the black (blue) contours denote the case of purely axial (equal axial and vector) couplings between the $Z'$ and SM fermions. For light $Z'$ masses, the cross section is dominated by the $s$-wave contribution, leading to an unsuppressed low-velocity cross section. For $m_{Z'} \gsim 2 m_{\chi}$, in contrast, the annihilation cross section at the temperature of thermal freeze-out is dominated by the $p$-wave contribution, and the low-velocity cross section is suppressed. The results shown here are for the case of dark matter with a 35 GeV mass, which provides a good fit to the gamma-ray excess for annihilations to $b\bar{b}$. 

The dotted contours shown in the lower portion of Fig.~\ref{majorana} represent the constraints on this model from direct searches, as derived from the results of the LUX experiment.  In the case of a $Z'$ with purely axial couplings to SM fermions, these constraints rule out $m_{Z'} \lsim 50$ GeV in this scenario. If there is any significant vector coupling, this leads to a large spin-independent scattering cross section that, despite the velocity suppression shown in Eq.~\ref{elasticmaj}, is ruled out by LUX. 

Taking the LUX constraint along with the requirement that $\sigma v (v \rightarrow 0) \gsim 4 \times 10^{27}$ cm$^3$/s (to generate the gamma-ray excess), we find that most of the parameter space is excluded, leaving only the case in which $m_{Z'}\simeq 50$ GeV. If we were to break the assumption that the $Z'$ couples equally to all SM fermions, however, much more of the parameter space could become viable.

\subsection{A Third Generation $Z'$}

In the previous two subsections, we assumed that the $Z'$ had equal (democratic) couplings to all SM fermions. The constraints from direct detection experiments, however, are driven in large part by the couplings of the $Z'$ to light quarks. One can relax such constraints (as well as constraints from accelerators) by considering dark matter that is mediated by a $Z'$ with suppressed couplings to light fermions.  A wide range of scenarios have been proposed in which a $Z'$ couples only to, or preferentially to, the third generation~\cite{Andrianov:1998hx}, including topcolor models~\cite{Hill:1994hp}, non-universal extended technicolor models~\cite{Chivukula:2002ry}, string inspired models~\cite{delAguila:1986iw}, and Kaluza-Klein models in which the fermion families are spatially separated~\cite{Belanger:2007dx}. With this motivation in mind, we will consider dark matter in the form of a Dirac fermion and a $Z'$ that couples to the SM only through the third generation:
\begin{equation}
\mathcal{L} \supset \left[ \bar{\chi} \gamma^\mu ( g_{\chi  v } + g_{\chi a} \gamma^5 ) \chi + \bar{f} \gamma^\mu ( g_{f  v } + g_{f a} \gamma^5 ) f \right] Z'_\mu, \\
\end{equation}
where $f=t, b, \tau, \nu_{\tau}$. The dark matter annihilation cross section (for each third generation fermion species) in this case is given by:
\begin{eqnarray}
\sigma &=&  \bigg[{\rm Eq}.~\ref{one} \bigg] \\
&+&\frac{1}{4 \pi  s \bL \left(s-m_{Z'} ^2\right)^2  + m_{Z'}^2 \Gamma_{Z'}^2 \bR } \sqrt{\frac{1-4 m_f^2/s}{1-4 m_{\chi }^2/s}} \, g^2_{fv} \nonumber \\
&\times& \Bigg[(2 m_f^2+s) \Big(g_{\chi a}^2 (s-4 m_{\chi }^2)+g_{\chi v}^2 (2 m_{\chi }^2+s)\Big) \Bigg]. \nonumber 
\end{eqnarray}

\begin{widetext}

Expanding in powers of $v^2$ (for each third generation fermion species):
\begin{eqnarray}
\sigma v &\approx&  \bigg[{\rm Eq}.~\ref{two} \bigg]+ \frac{3 \sqrt{1-m_f^2/m_\chi^2}}{2 \pi  \left(m_{Z'}^2-4 m_{\chi }^2\right)^2} \, g_{fv}^2 g_{\chi v}^2 \left(m_f^2+2 m_{\chi }^2\right) \nonumber \\
&-& \frac{v^2}{16 \pi  m_{\chi }^2 \sqrt{1-m_f^2/m_{\chi }^2} \left(4 m_{\chi }^2-m_{Z'}^2\right)^3} 
\, g_{fv}^2 \bigg\{ g_{\chi v}^2 \Big(8 m_{\chi }^4 \left(m_{Z'}^2-4
   m_f^2\right)-4 m_f^2 m_{\chi }^2 \left(17 m_f^2+m_{Z'}^2\right)
   \\&+&5 m_f^4 m_{Z'}^2+64 m_{\chi }^6\Big)-4 g_{\chi a}^2 \left(m_f^2 m_{\chi
   }^2+m_f^4-2 m_{\chi }^4\right) \left(m_{Z'}^2-4 m_{\chi }^2\right)\bigg\} \nonumber.
\end{eqnarray}
\end{widetext}

\begin{figure*}
\includegraphics[width=2.32in]{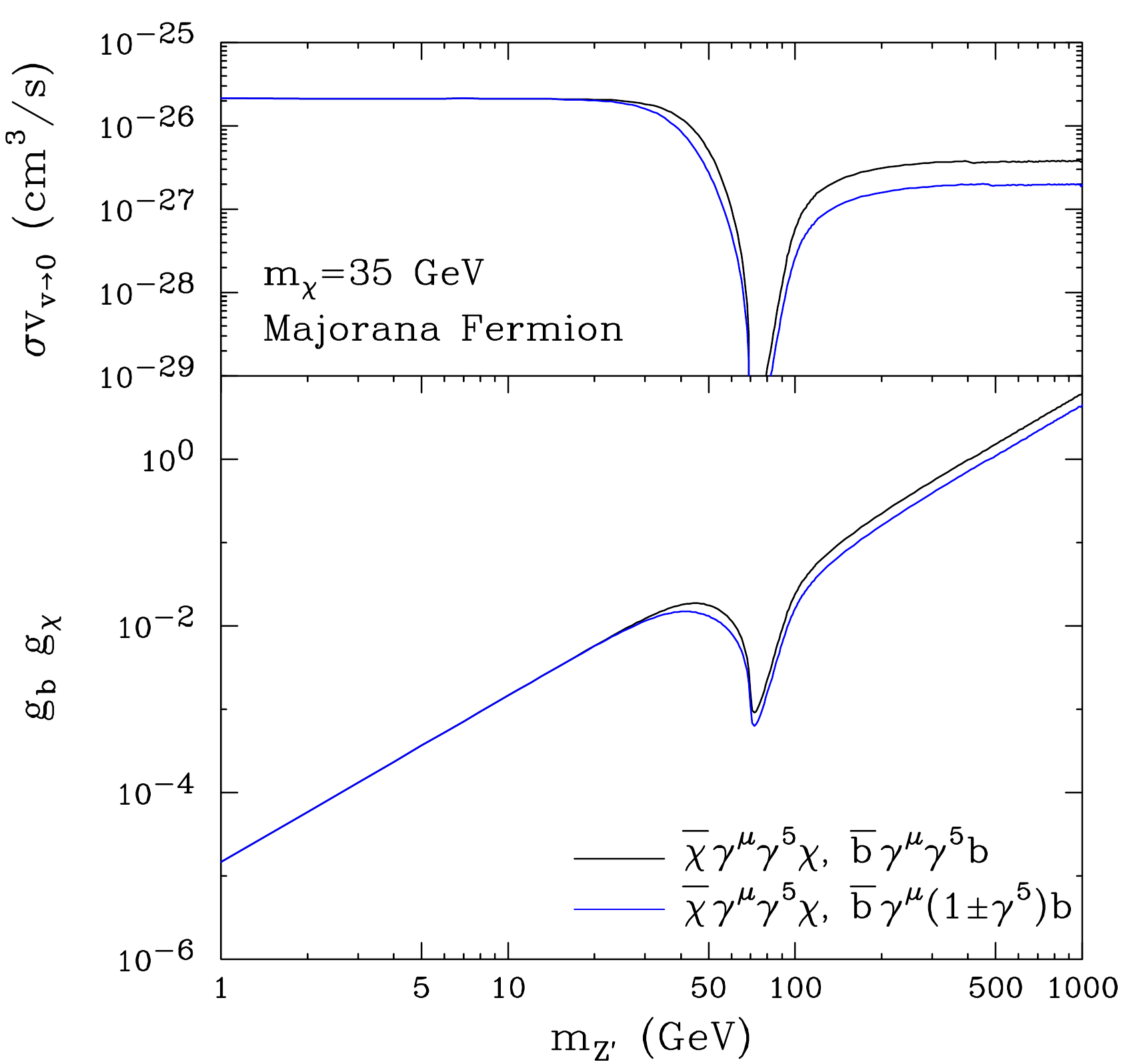} 
\includegraphics[width=2.32in]{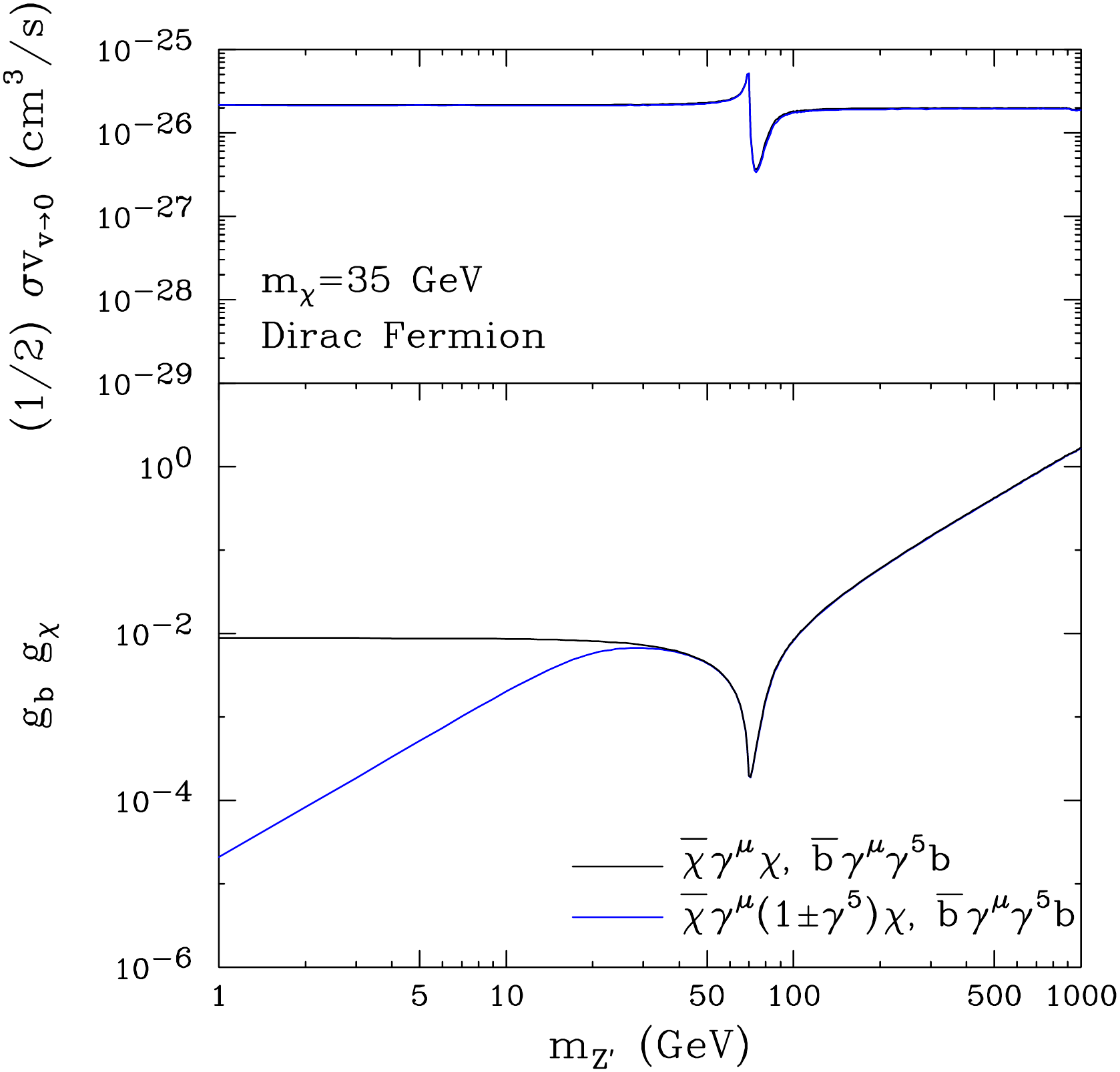} 
\includegraphics[width=2.32in]{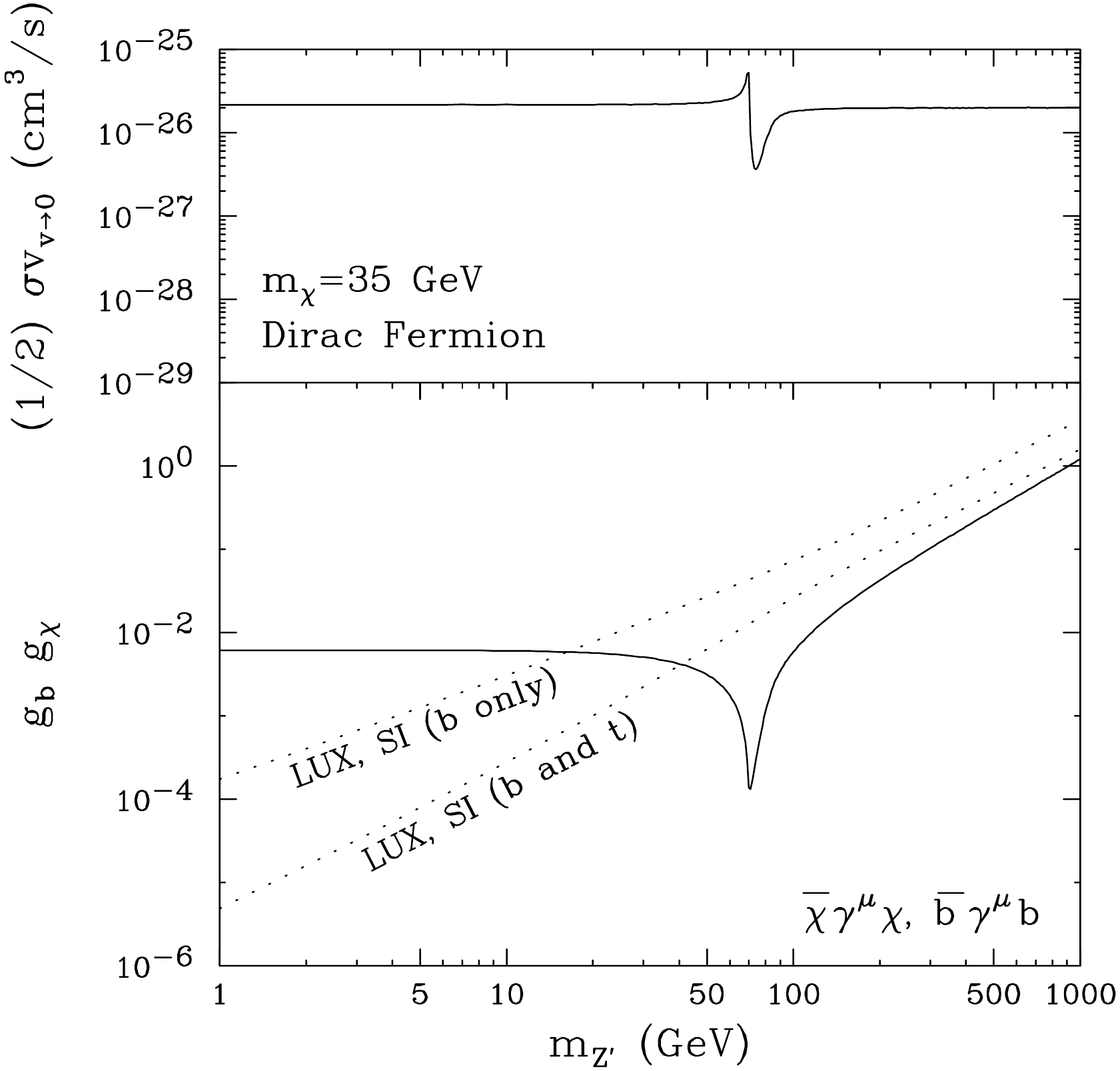} 
\caption{Results for dark matter annihilating through the exchange of a $Z'$ with couplings only to third generation quarks. Notation is the same as in previous figures.}
\label{3gen}
\end{figure*}

In these expressions, $[{\rm Eq}.~\ref{one}]$ and $[{\rm Eq}.\ref{two}]$ denote the cross sections as provided in the respective equations, for a given fermion species (as opposed to the sum over all fermion final states).

In the left and center frames of Fig.~\ref{3gen}, we show the results for scenarios described in the previous two sub-sections, as modified such that the $Z'$ couples only to the dark matter and $b$-quarks (the results would be almost identical if the couplings were assumed to extend to the rest of the third generation).  Although much of the phenomenology remains unchanged, two important differences are immediately apparent. First, direct detection constraints (and prospects) all but disappear; without couplings to first and second generation quarks, the elastic scattering cross section with nuclei become sufficiently small to evade all planned or proposed direct detection experiments. Second, in the case of Majorana dark matter (left frame), the low-velocity annihilation cross section is larger than in the case of democratic couplings, and is thus more capable of accounting for the observed normalization of the Galactic Center gamma-ray excess. In the right frame of the same figure, we consider the possibility of a $Z'$ with vector couplings to both the (Dirac) dark matter and to third generation quarks. In this case, the elastic scattering cross section with nuclei is generated through a bottom loop coupled to a photon~\cite{Agrawal:2011ze} (loops with two gluons do not contribute~\cite{Kaplan:1988ku,Ji:2006vx}).

\section{Collider Constraints on a Dark Matter Mediating $Z'$}
\label{collidersec}

In this section, we discuss the constraints from accelerator experiments on a $Z'$ that mediates the annihilations of dark matter. Although searches for dark matter particles through their jet-plus-missing energy signature at hadron colliders ({\it ie.} mono-jets)~\cite{ATLAS:2012ky,ATLAS:2012zim,Chatrchyan:2012me,Khachatryan:2014rra,Aad:2014vea} are potentially very interesting, the accelerator constraints on the $Z'$ itself are generally more restrictive. For this reason, we focus on $Z'$ constraints. 

\begin{figure*}
\includegraphics[width=4.6in]{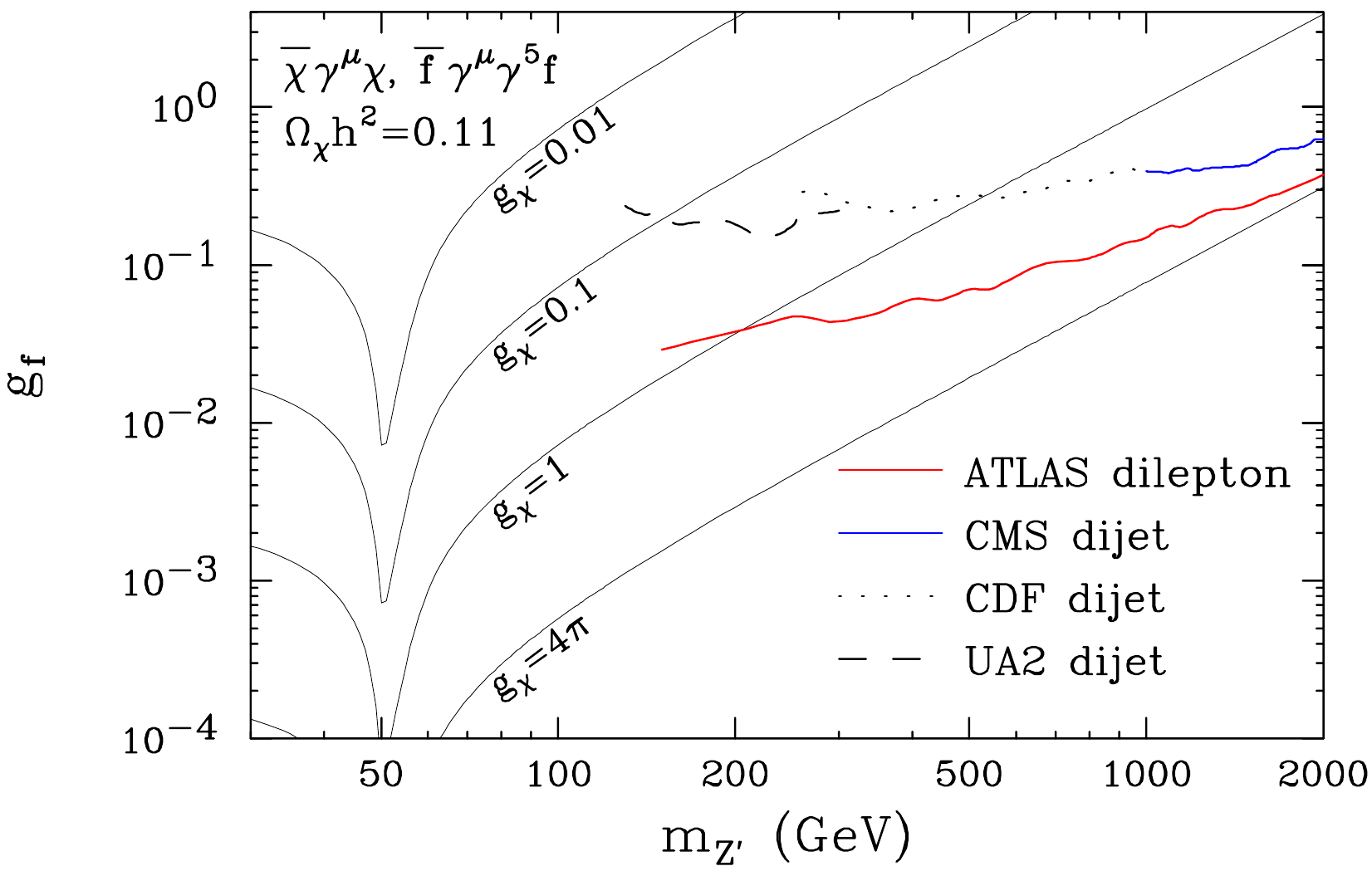} 
\caption{Constraints from dilepton resonance searches at ATLAS (solid red), and dijet resonance searches at UA2~\cite{Alitti:1993pn} (dashed), CDF~\cite{Aaltonen:2008dn} (dotted), and CMS~\cite{CMS:2012yf,Chatrchyan:2013qha} (solid blue) on the couplings a $Z'$ that mediates the annihilations of the dark matter. Throughout the plane, $g_{\chi}$ has been set such that the thermal relic abundance is equal to the cosmological dark matter density (black solid lines represents contours of fixed $g_{\chi}$). For a $Z'$ with equal couplings to all Standard Model fermions, the dilepton constraints are very stringent, generally forcing us toward scenarios in which $g_{\chi} \gg g_f$. If the $Z'$ is leptophobic (coupling less strongly to leptons than to quarks), only the dijet constraints apply, and $g_{\chi}$ can be within a factor of a few of $g_q$ for $m_{Z'} \lsim 500$ GeV.}
\label{collider}
\end{figure*}

By searching for peaks in the invariant mass distribution of dilepton events, as predicted to result from the process $pp \rightarrow Z' X \rightarrow l^+ l^- X$ (where $l=e$, $\mu$), the ATLAS~\cite{Aad:2014cka} and CMS~\cite{Chatrchyan:2012oaa} Collaborations have placed very stringent constraints on a $Z'$ with unsuppressed couplings to both light quarks and leptons. The rate for such events is determined by the product of the $Z'$ production cross section and its branching fraction to electrons or muons, which are determined in large part by its couplings to light quarks and leptons, respectively. In Fig.~\ref{collider}, we have translated the constraint of Ref.~\cite{Aad:2014cka} into the $g_f$-$m_{Z'}$ plane, for the case of a $Z'$ with purely axial and democratic couplings to SM fermions, and setting the value of $g_{\chi}$ to ensure a thermal relic abundance in agreement with the cosmological dark matter density. In this case, with equal couplings to all SM fermions, a $Z' $ heavier than $\sim$150 GeV is only viable as a mediator of dark matter annihilation if its effective coupling to the dark matter is much larger than that to SM fermions, by a factor of $\sim$\,20 or more.  Unless near resonance ($m_{Z'} \simeq 2 m_{\chi}$), similar conclusions are reached for lower masses, as a result of constraints on the leptonic coupling of a $Z'$ from LEP II, which limit $g_{e} \lsim 10^{-2}$ for $m_{Z'} \lsim 209$ GeV~\cite{lep:2003ih}.\footnote{The $Z'$ constraints from LEP II are more stringent than those from the LHC only if the couplings are highly hadrophobic ({\it ie.} leptophillic) or if $m_{Z' } \lsim 200$ GeV.} Although the charge assignments of a given $U(1)'$ are generally expected to all be of the same order of magnitude,  a large hierarchy of coupling strengths could result if the $Z'$ obtains an effective coupling to the SM through mixing.

A $Z'$ with suppressed couplings to electrons and muons can be much more difficult to constrain using either $e^+ e^-$ or hadron colliders. In the case of a leptophobic $Z'$, the most stringent constraints are generally derived from from searches for a dijet resonance at hadron colliders, via the process $q\bar{q} \to Z' \to q'\bar{q}'$. These constraints are shown in Fig.~\ref{collider}, including limits from UA2~\cite{Alitti:1993pn}, CDF~\cite{Aaltonen:2008dn}, and CMS at both 7 TeV~\cite{CMS:2012yf} and 8 TeV~\cite{Chatrchyan:2013qha}. The sensitivity of such searches is ultimately limited by the large and uncertain QCD background, which swamps any would-be resonance arising from a $Z'$ with couplings smaller than approximately electroweak-strength.

In addition to these more direct accelerator constraints, there are also a number of indirect and low energy measurements that restrict the mass and couplings of a $Z'$. In particular, kinetic mixing between the $Z'$ and the SM $Z$ could shift the mass of the $Z$ from its predicted value, contributing to the $T$ parameter~\cite{Peskin:1991sw}.  Precision measurements of $m_Z$ (as well as other electroweak measurements) can thus strongly constrain the degree of mixing between the $Z$ and any $Z'$~\cite{Langacker:1991pg,Umeda:1998nq}. 

If the couplings of the $Z'$ to the SM quarks are generation dependent, tree-level flavor-changing neutral current processes will be generated~\cite{Langacker:2000ju}, strongly constrained by meson mixing measurements~\cite{Langacker:2000ju, He:2004it, Gupta:2010wt}.  To evade the most stringent of such constraint, we assume throughout this study that the $Z'$ couples identically to the first two generations of the SM. And although quarkonium decays can be restrictive for $m_{Z'}\lsim$10 GeV~\cite{Dobrescu:2014fca}, heavier $Z'$s are not strongly constrained by such observations.

As we will discuss in the following section, additional fermions are generally required in order to cancel gauge anomalies in theories with a $U(1)'$. In addition to directly searching for a $Z'$, colliders can also be sensitive to these exotic fermions. The requirement of perturbativity translates into an upper limit on the mass of such exotics, $m_{f} \lsim 5.4 \, {\rm TeV} \times (m_{Z'}/100 \, {\rm GeV}) (0.1/g_{Z'}) (1/z_{\varphi})$, where $z_{\varphi}$ is the charge of the Higgs field responsible for breaking the $U(1)'$.  The prospects for the discovery of such particles at the LHC seem encouraging, in particular for any exotics with QCD color.

\section{Anomaly-Free Models}
\label{anomaliessec}


In order to ensure gauge invariance of the full quantum field theory, all anomalies that arise due to triangle diagrams with gauge bosons as external lines must cancel. In this section, we will discuss what additional fermionic particle content must be introduced into a theory with an additional $U(1)'$ if it to be anomaly-free. 

Before proceeding, we need to establish some notation. So far in this paper, we have described the interactions of a $Z'$ in terms of its effective vector and axial couplings, $g_{fv}$ and $g_{fa}$. Assuming that the $Z'$ corresponds to a single $U(1)'$ (as opposed to being a mixture of different gauge eigenstates), we can write these effective couplings in terms of $U(1)'$ gauge charges:
\begin{eqnarray}
g_{fv} &=& g_{Z'} \bigg(\frac{z_{f_L}+z_{f_R}}{2}\bigg) \\
g_{fa} &=& - g_{Z'} \bigg(\frac{z_{f_L}-z_{f_R}}{2}\bigg), \nonumber
\end{eqnarray}
where $z_{f_L}$ and $z_{f_R}$ are the gauge charges of left and right-handed fermions, and $g_{Z'}$ is the gauge coupling of the $U(1)'$.

The anomalies resulting from triangle diagrams with two gluons and a $Z'$ or two $SU(2)_W$ gauge bosons and a $Z'$ are cancelled if the following conditions are met:
\begin{eqnarray}
\sum_{f_L \in 3,3^*} n_2 z_{f_L}  - \sum_{f_R \in 3,3^*} n_2 z_{f_R}   = 0 
\label{anomaly1}
\end{eqnarray}
and
\begin{eqnarray}
\sum_{f_L \in 2} n_c z_{f_L} - \sum_{f_R \in 2} n_c z_{f_R}   = 0, 
\end{eqnarray}
where the sums extend over all left- and right-handed fermions charged under $SU(3)$ and $SU(2)$, respectively. The counting factor $n_2$ is equal to 2 (1, 3) for all fermions in an $SU(2)$ doublet (singlet, triplet) and $n_c$ is 3 (1, 8) for fermions in an $SU(3)$ triplet (singlet, octet). 

\begin{table*}[htdp]
\begin{center}
\begin{tabular}{|c|c|c|c||c|c|c|c|c|c|}
\hline  &  {\bf \emph{$SU(3)_C$}} &  {\bf \emph{$SU(2)_W$}} & {\bf \emph{$U(1)_Y$}} & Axial A  &  Axial B  &  Leptophobic A & Leptophobic B & Leptophobic C & Axial-Leptophobic\\ 
\hline \hline 
$q_L$ & 3 & 2  &   1/3 &  1/3 & 1/3 & 1/3 & 1/3 & 2/3 & 1/3\\ \hline
$u_R$ & 3 & 1 &  4/3&  -1/3 & -1/3 & 1/3 & 1/3 & 2/3 & -1/3\\ \hline
$d_R$ & 3 & 1  &  -2/3&  -1/3 & -1/3 & 1/3 & 1/3 & 2/3 & -1/3\\ \hline
$l_L$ & 1 & 2 &  -1 & 1/3 & -1/3 & 0 & 0 & 0 & 0\\ \hline
$e_R$ & 1 & 1 &    -2 & -1/3 & -2/3 & 0 & 0 & 0& 0\\ 
\hline \hline
$\nu_R$ & 1 & 1  &   0  & -1/3 & -- & -1 & -3 & -- & -5/3\\ \hline
$\nu'_R$ & 1 & 1  &  0   & -4/3  &  -- & -- & 2 & -- & -- \\ \hline
$\chi_L$ & 1 & 1 &    0 & -- & 1/3 & -- & -- & 1& --\\ \hline
$\chi_R$ & 1 & 1 &  0 & -- & -4/3 & -- &-- & -1& --\\ 
\hline \hline
$\psi^d_L$ & 3 & 1 &  -2/3 & -2/3  & 2/3 & -- &--&-- & -1/3\\ \hline
$\psi^d_R$ & 3 & 1 & -2/3 & 2/3  &  -2/3 & -- &--& -- & 1\\  \hline 
$\psi^l_L$ & 1 & 2    & -1  &   -2/3  &  2/3 & -1 &2 & -1 & -1\\ \hline
 $\psi^l_R$ & 1 & 2  & -1 & 2/3  &  1/3 & -- &3 & 1 & --\\  \hline
$\psi^e_L$ & 1 & 1    & -2  &   --  &  -- & -1 &3 & 1& --\\ \hline
 $\psi^e_R$ & 1 & 1  & -2 & --  &  -- & -- &2 & -1 & -1/3\\ 
\hline 
\end{tabular}
\end{center}
\caption{Charge assignments for Standard Model fermions and exotic fermions in selected $U(1)'$ scenarios.}
\label{anomalies}
\end{table*}%

Moving on, the cancellation of anomalies resulting from diagrams with a combination of two hypercharge gauge bosons and a $Z'$ requires:
\begin{eqnarray}
\sum_{f_L} n_c n_2 Y^2_{f_L} z_{f_L}  - \sum_{f_R} n_c n_2 Y^2_{f_R} z_{f_R}   = 0,
\end{eqnarray}
where $Y_f$ is the fermion's hypercharge.
And similarly for the diagram with two $Z'$s and one hypercharge gauge boson:
\begin{eqnarray}
\sum_{f_L} n_c n_2 Y_{f_L} z^2_{f_L}  - \sum_{f_R} n_c n_2 Y_{f_R} z^2_{f_R}   = 0. 
\end{eqnarray}
The cancellation of the self-induced anomaly from a triangle with three $Z'$s requires:
\begin{eqnarray}
\sum_{f_L} n_c n_2 z^3_{f_L}  - \sum_{f_R} n_c n_2 z^3_{f_R}   = 0. 
\end{eqnarray}
And lastly, the cancellation of the anomaly resulting from a diagram with two gravitons and a $Z'$ requires:
\begin{eqnarray}
\sum_{f_L} n_c n_2  z_{f_L}  - \sum_{f_R} n_c n_2  z_{f_R}   = 0. 
\end{eqnarray}

If the only fermions in these loops are those of the SM, the only anomaly-free $U(1)'$ is $U(1)_{B-L}$ ({\it ie.} that charged under baryon number minus lepton number). All other charge assignments require the introduction of additional chiral fermions, known as exotics~\cite{Batra:2005rh,Appelquist:2002mw}. As $U(1)_{B-L}$ leads neither to purely axial nor leptophobic couplings, we are lead to consider models with additional fermions. A $(B-L)$ charge assignment limited to third generation fermions would be a phenomenologically viable option, however.

Following Secs.~\ref{axial} and~\ref{majoranasec}, we first consider a $Z'$ with purely axial couplings to light quarks, $z_{q_L} = - z_{u_R} = - z_{d_R}$. In this case, Eq.~\ref{anomaly1} (describing the $g-g-Z'$ anomaly) cannot be satisfied without introducing colored fermions beyond those contained within the SM.  In Table~\ref{anomalies}, we describe the particle content in examples of anomaly-free $Z'$ models with purely axial couplings to SM quarks. The first of these cases (labelled ``Axial A'') is the well known $U(1)_{\psi}$ model, as inspired by $E_6$ Grand Unification. In this model, all fermions belonging to the 10 or $\bar{5}$ representations of $SU(5)$ have the same charge under the new $U(1)'$. Anomalies are cancelled in this model by two right-handed neutrinos ($\nu_R$, $\nu'_R$), as well as the fermions $\psi^l$ and $\psi^d$, which are vector-like under the SM and possess the same SM charges as $l_L$ and $d_R$, respectively (each per generation). In this model, any of the right-handed neutrinos could serve as our dark matter candidate, assuming that it is stabilized by some unnamed symmetry. We also present in Table~\ref{anomalies} an alternative model with purely-axial couplings to the $Z'$ (labelled ``Axial B''). Here, instead of a pair of right-handed neutrinos, we include left- and right-handed components of the fermion $\chi$, which again can serve as our dark matter candidate.

Moving on to leptophobic scenarios, we consider three models with a $Z'$ that couples proportionally to baryon number (``Leptophobic A-C'')~\cite{Carone:1994aa,FileviezPerez:2010gw}. Although these models do not require any additional colored fermions to cancel anomalies, they do require $\psi^l$ and $\psi^e$, with the same SM charges as $l_L$ and $e_R$, respectively. The scenario ``Leptophobic B'' is particularly interesting, as the $Z'$ couples to the potential dark matter candidate $\nu_R$ nine times more strongly than it does to SM quarks, resulting in significantly relaxed constraints from hadron colliders.

\begin{figure}
\includegraphics[width=3.3in]{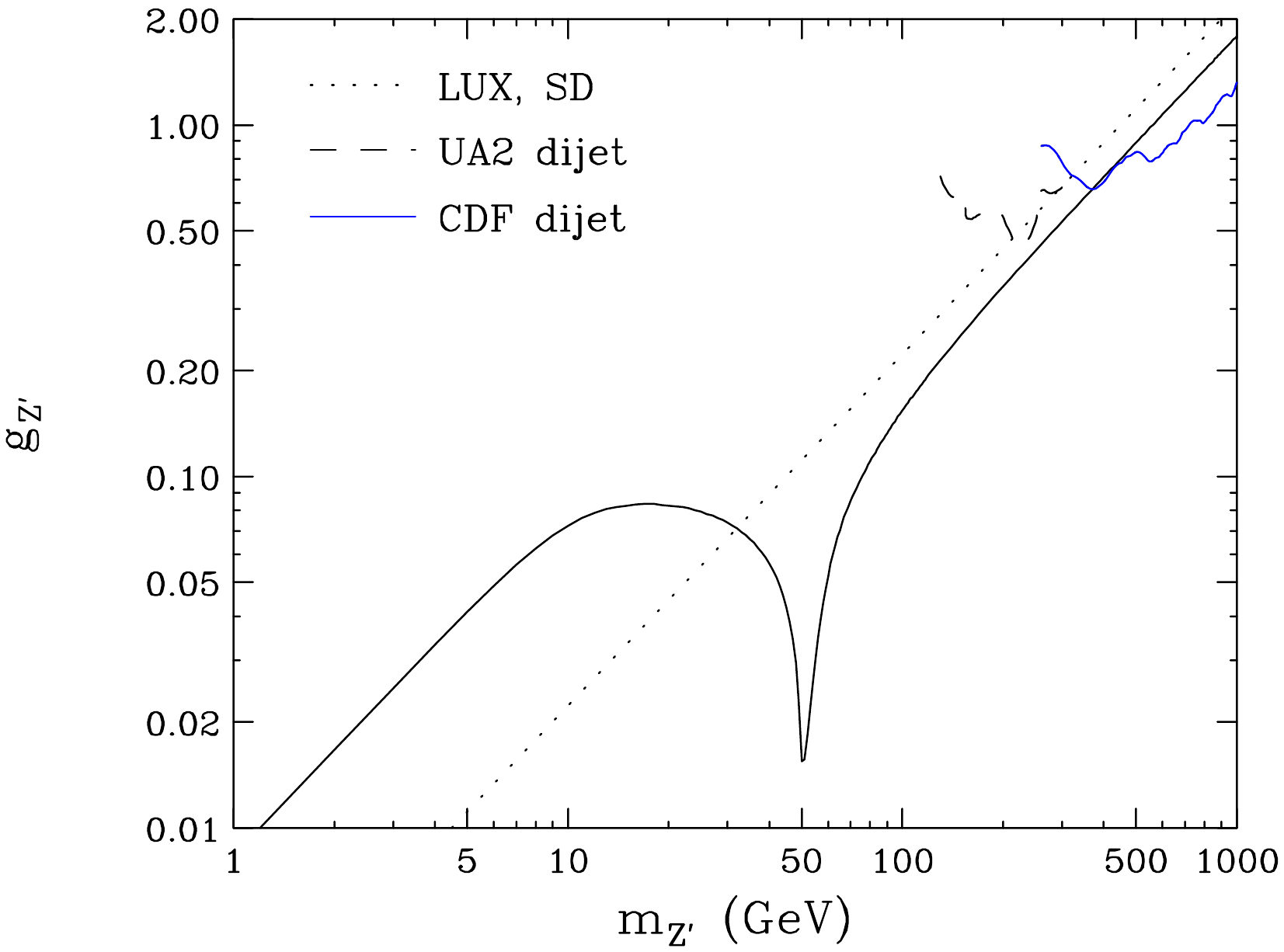} 
\caption{Collider (CDF~\cite{Aaltonen:2008dn}, UA2~\cite{Alitti:1993pn}) and direct detection (LUX~\cite{Akerib:2013tjd}) constraints on the axial and leptophobic $Z'$ model described in the rightmost column of Table~\ref{anomalies}. The solid black contour is the region of the plane in which the thermal relic abundance is equal to the cosmological dark matter density. For $m_{Z'} \simeq 30-400$ GeV, this model provides a viable explanation for the Galactic Center gamma-ray excess.}
\label{axialleptophobic}
\end{figure}

To the best of the author's knowledge, there are no well known examples in the literature of $U(1)'$ models that are both axially coupled to quarks and leptophobic.  Such a model is not difficult to construct, however, and we give an example in the final column of Table~\ref{anomalies} (``Axial-Leptophobic"). In Fig.~\ref{axialleptophobic} we plot the constraints on the parameter space of this axial and leptophobic model. For a wide range of $Z'$ masses ($m_{Z'} \simeq 30-400$ GeV) this model is compatible with all direct detection and accelerator bounds.

In addition to axial and leptophobic models, there are many viable scenarios in which the $Z'$ couples dominantly to the third generation. For example, the models ``Axial A'' and ``Axial B'', if applied only to the third generation, would lead to phenomenology similar to that shown in the middle frame of Fig.~\ref{3gen}.



\section{Prospects for Future Direct Detection Experiments}
\label{future}

We would like to emphasize the importance of upcoming results from direct detection experiments in testing the range of models considered in this study.  Assuming that the $Z'$ couples to light quarks, the XENON1T experiment (and possibly LUX) should be sensitive to most of the scenarios presented here; the only exceptions being a $Z'$ with purely vector couplings to the dark matter and purely axial couplings to the SM, or a $Z'$ near resonance ($m_{Z'} \simeq 2 m_{\chi}$).  Alternatively, if the $Z'$ only couples to third generation fermions, direct detection experiments will soon become sensitive to any vector interactions that couple the dark matter to $b$ or $t$ quarks. Axial couplings to the third generation will remain beyond the reach of such experiments for the foreseeable future. 

We note that similar conclusions are reached when considering other classes of particle physics models capable of accounting for the Galactic Center gamma-ray excess. For example, if dark matter annihilations are mediated by a pseudoscalar (such as in two Higgs doublet models, etc.), the scalars that are also found in such models are generally predicted to generate elastic scattering cross sections that are within the reach of upcoming direct detection experiments~\cite{Ipek:2014gua,Cheung:2014lqa,Cerdeno:2014cda}. Similarly, models in which the dark matter annihilates to SM fermions through $t$-channel diagrams predict elastic scattering cross sections that are not far below the present bounds from LUX~\cite{Agrawal:2014una,Berlin:2014tja}.  

In light of these considerations, if no positive signal is reported from LUX or XENON1T, hidden sector models will become increasingly attractive. Models with an additional $U(1)'$ naturally lead to such phenomenology when $m_{Z'} < m_{\chi}$~\cite{Hooper:2012cw,Berlin:2014pya,Cline:2014dwa,Abdullah:2014lla}. Although we have not focused on this case here, such models are simple to construct and can quite easily evade constraints from both direct detection and collider experiments.

\section{Summary and Conclusions}
\label{conclusion}

Additional massive neutral spin-one particles appear within a wide variety of well motivated scenarios for physics beyond the Standard Model. In this study, we have considered models in which the dark matter interacts with the Standard Model through the exchange of a such a $Z'$, with the intention of identifying scenarios that are capable of producing the gamma-ray excess observed from the region surrounding the Galactic Center. After taking into account all constraints from colliders (LHC, Tevatron, LEP II, UA2) and direct detection experiments (most notably, LUX), we find three viable classes of such models: 

\begin{itemize}
\item{A $Z'$ with axial couplings to Standard Model quarks and suppressed couplings to leptons can fulfill each of these requirements. We have identified an example of an anomaly-free $U(1)'$ that leads to an axial and leptophobic $Z'$ which mediates the annihilations of a right-handed neutrino that serves as our dark matter candidate. For $m_{Z'}\simeq 30-400$ GeV, this models is consistent with all collider and direct detection constraints.}

\item{A $Z'$ that couples preferentially to third generation fermions can mediate the annihilations of dark matter responsible for the Galactic Center excess, while being largely unconstrained by direct detection and collider experiments.}

\item{A quite weakly coupled $Z'$ with a mass near the dark matter annihilation resonance ($m_{Z'} \simeq 2 m_{\rm DM}$) can evade all current constraints, even without suppressed couplings to leptons.}

\end{itemize}  

Many of these scenarios are within the anticipated reach of upcoming direct detection experiments, including LUX and XENON1T. A null result from this collection of experiments would significantly reduce the range of dark matter models that could potentially account for the Galactic Center gamma-ray excess.

\bigskip \bigskip \bigskip

{\it Acknowledgements}: We would like to thank Sam McDermott, Asher Berlin, Claudia Frugiuele, Prateek Agrawal, and Yann Manbrini for helpful interactions. This work has been supported by the US Department of Energy under contract DE-FG02-13ER41958. Fermilab is operated by Fermi Research Alliance, LLC, under Contract No.~DE-AC02-07CH11359 with the US Department of Energy.

\bibliography{zprimegc}

\end{document}